\documentclass[]{tMOP2e}

\begin{document}

\title{New factorization algorithm based on a continuous representation of truncated Gauss sums}

\author{Vincenzo Tamma$^{a}$$^{b}$$^{\ast}$\thanks{$^\ast$Corresponding author. Email: tammav1@umbc.edu
\vspace{6pt}}, Heyi Zhang$^{a}$, Xuehua He$^{a}$, Augusto Garuccio$^{b}$ and Yanhua Shih$^{a}$\\\vspace{6pt}  $^{a}${\em{ Department of Physics, University of Maryland,
Baltimore County, Baltimore, Maryland 21250, USA}};
$^{b}${\em{Dipartimento Interateneo di Fisica, Universit\`{a} degli Studi di
Bari, 70100 Bari, Italy}}}
\maketitle

\begin{abstract}
 In this paper, we will describe a new factorization algorithm based on the continuous representation of Gauss sums, generalizable to orders $j>2$. Such an algorithm allows, for the first time, to find all the factors of a number $N$ in a single run without precalculating the ratio $N/l$, where $l$ are all the possible trial factors.
Continuous truncated exponential sums turn out to be a powerful tool for distinguishing factors from non factors (we also suggest, with regard to this topic, to read an interesting paper of S. W\"{o}lk et al. published in this issue\cite{wolk}) and factorizing different numbers at the same time. We will also describe two possible $M$-path optical interferometers, which can be used to experimentally realize this algorithm: a liquid crystal grating and a generalized symmetric Michelson interferometer.
\end{abstract}

\bigskip

\section{Introduction}

The factorization of large numbers is one of the problems for which
classical computers need an exponential number of resources. Quantum computers, instead, need a polynomial number of resources. In fact, in $1994$, Shor showed a quantum algorithm, which is able, in principle, to factorize a number $N$, using a number of resources
and a number of runs polynomial in $log N$, exploiting
quantum entanglement\cite{shor}. Unfortunately, $15$ is the largest factorized number, so far, using Shor's algorithm. For this reason, a great research interest has been directed towards the implementation of classical analogue computers able to factorize larger numbers, even if they need an exponential number of resources.

\subsection{Proposals of factorization with classical interference}

An interesting classical scheme was proposed, in $1996$, by Clauser and Dowling\cite{dowling}. The setup is given by a Young's $N$-slit interferometer with period $a$, where $N$ is the number to factorize. The intensity peaks, of the diffraction pattern, at distance $R$ from the $N$-slit arrangement, have same amplitude only if the quantity $n=\lambda \frac{R}{(a^2)}$ is a factor of $N$, where $\lambda$ is the wavelength of the incident radiation. The drawback of this scheme is that the number of interfering paths must equal the number $N$ to factorize. Consequently, the number of resources associated with the slits in the interferometer increase substantially as $N$ increases and the setup needs to be modified depending on the number $N$ to factorize. Moreover, it is necessary to run the procedure for each trial factor.

Later Summhammer proposed\cite{summhammer} a different interferometry scheme for the
realization of truncated Fourier-like sums with a truncation parameter given by the particular trial factor $l$. Unfortunately Summhammer's approach also presents several drawbacks. First,  the ratio between $N$ and any trial factor $l$ is calculated before the experiment is actually performed, in order to fix the phases of the phase shifters in the Mach Zehender interferometers.
Second, Summhammer's sums are obtained by summing the intensities of the outcoming light from different interferometers. This unfortunately means that there are less interfering terms and consequently more relative background noise, with  respect to the output intensity pattern of a single $l$-path interferometer, which is instead able to reproduce all the interfering terms in the modulo square of a Fourier sum with the same number $l$ of terms.
Third, in Sumhammer's approach, in order to check different trial factors at the same time, it is necessary to combine a number of Mach-Zehender interferometers and a number of detectors equal to the number of trial factors to check.

\subsection{Factorization with Gauss sums}

A more recent approach to factorization proposed by W. Schleich
exploits the periodic properties of truncated exponential sums
\cite{merkel2,merkel3,merkel,ghost,expsum} of order $j$:
\begin{eqnarray}\label{exp}
{\cal A}_N^{(M,j)}(\ell) = \frac{1}{M} \sum_{m=1}^M \exp\left[
2\pi i\, (m-1)^j\frac{N}{\ell}\right],
\end{eqnarray}
where $M$ is the number of phase terms in the sum ($M$ is called
the truncation parameter), $N$ is the number to be factored, and $j$
and $l$ are positive integers, with $j>1$ and $1\leq\ell \leq
\sqrt{N}$. For $j=2$, the truncated exponential
 sum reduces to a truncated Gauss sum \cite{merkel}.  If $\ell$ is a factor of $N$, all the
terms interfere constructively and the  modulo squared of the truncated exponential sum
assumes its maximum value, i.e. $1$. On the other hand, if $\ell$ is
not a factor of $N$,  the modulo squared of the truncated exponential sum assumes a value
less than one, because of the destructive interference caused by the
rapid oscillation of the phase terms of order $j$ in Eq.
(\ref{exp}).  It turns out that only a relatively few number of terms, compared to $N$, is necessary in order to discriminate factors
from non factors \cite{expsum}. Consequently, Schleich's approach substantially reduces the number of necessary interfering terms in the factorization procedure with respect to the schemes proposed by Clauser and Dowling and Summhammer.

The main goal of the Gauss sums approach is
the implementation of an analogue computer, instead of a digital computer, able to
factorize numbers, through the implementation of exponential
sums.

\subsubsection{Drawbacks in the past realizations of the Gauss sums procedure}

Gauss sums  have been
reproduced experimentally \cite{nmr1,nmr2,cold,pulses,bec,randomweber,randompeng}, demonstrating that the Gauss sum factorization procedure proposed by Schleich is feasible.

In all these realizations, Gauss sums are realized by storing, in a suitable physical system, electromagnetic quadratic phases proportional to the global ratio $N/l$. Consequently, as pointed out by Jones\cite{jones}, in order to determine such phases and actually perform the experiment, is necessary to evaluate the ratio $N/l$ before the experiment is run. Another important drawback is the fact that it is necessary to run the experiment for each trial factor and to change the experimental setup for each number we want to factorize.

\section{Goals in a realistic factorization approach}

We have described all drawbacks present in the past proposed schemes for factorization. Our goal consists of achieving a realistic factorization procedure, which fulfills, at the same time, three important goals:
a)	no calculation of the ratio between $N$ and $l$ before the experiment is run;
b)	determination of the factors in only a single run of the experiment;
c) use of the same experimental setup for factorizing different numbers.

\section{Realistic factorization with exponential sums of a definite number $N$}
We will now point out how it is possible to achieve the first two goals, stated before, for a realistic factorization  of a definite number $N$ with the exponential sums procedure. First of all, we need to introduce a suitable physical system, which exploits interference in order to reproduce truncated exponential sums. The first goal can be achieved if such a system can provide  two independent physical parameters $p_N$ and $p_l$, proportional to $N$ and $l$, respectively, which can be varied independently in the experiment, preventing us from knowing in advance their ratio. In particular, the following correspondences need to be satisfied:
\begin{eqnarray}\label{cond1}
p_{N} \equiv  N u_N,
\end{eqnarray}
\begin{eqnarray}\label{cond2}
p_{l} \equiv  l u_l,
\end{eqnarray}
where $u_N$ and $u_l$ are the unit of measurement associated to the two physical observables $p_N$ and $p_l$, respectively.
The modulo squared of  the truncated exponential sum in Eq. \ref{exp} is given, apart from a constant, by the intensity pattern, as a function of the two parameters $p_{N}$ and $ p_{l}$, associated to the interference process in the physical system:
\begin{eqnarray}\label{expphys}
|{\cal A}_{p_N}^{(M,j)}(p_l)|^{2} = |\frac{1}{M} \sum_{m=1}^{M} \exp\left[
2\pi  i\frac{p_{N,m}}{p_{l}}\right]|^{2},
\end{eqnarray}
where we have defined a global parameter
\begin{eqnarray}\label{cond3}
p_{N,m}\doteq\frac{u_l}{u_N}p_m p_N,
\end{eqnarray}
 which include the exponential phase terms $p_m \equiv (m-1)^j$, for each $m=1,...,M$.
 The constant term $\frac{u_l}{u_N}$, in Eq. \ref{cond3}, makes sure that the phase terms in Eq. \ref{expphys} are dimensionless.
 From Eq. \ref{expphys}, it is clear that the two parameters $p_{N,m}\propto N$ and $p_{l}\propto l$ represent the two independent input conditions of a general experimental setup for factorization and only, through the actual result of the experiment, i.e. the reproduced exponential sum, we can infer about their ratio.
 On the other hand, in order to reproduce the exponential sums for each possible trial factor in a single run, the experimental result needs to contain information about not only one, but all the values of the parameter $p_{l}\propto l$, associated with all the possible trial factors.

\begin{figure}
\begin{center}
\resizebox*{12cm}{!}{\includegraphics{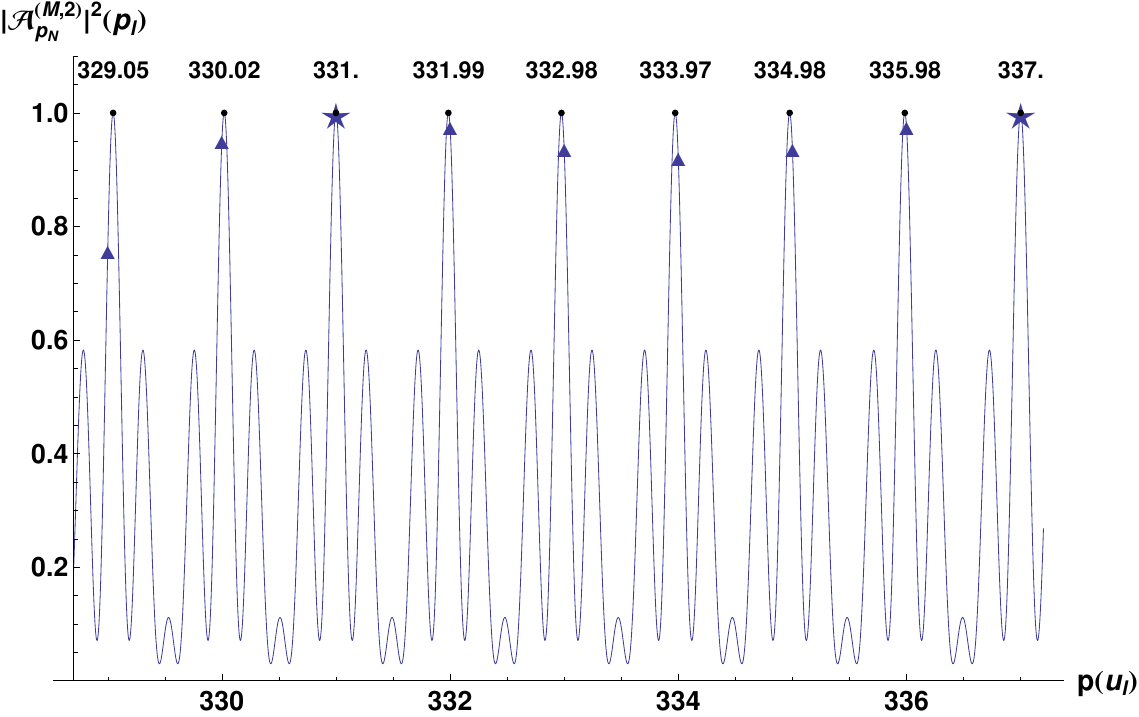}}%
\hfill
\resizebox*{12cm}{!}{\includegraphics{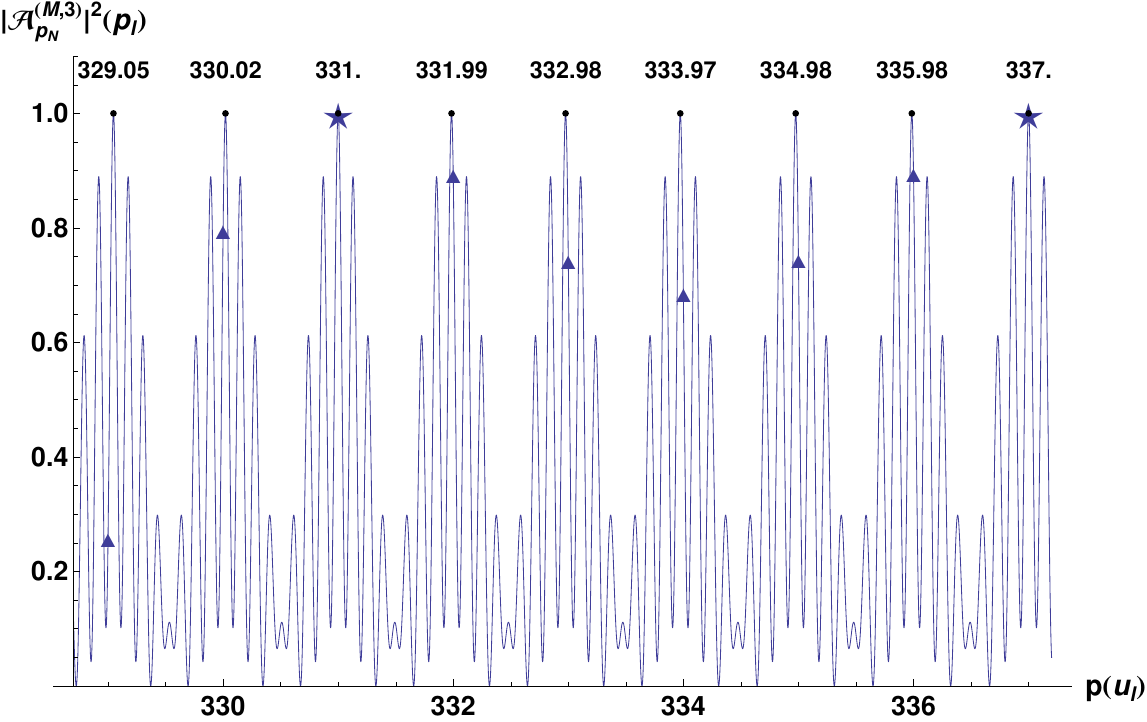}}%
\caption{Continuous representation of a truncated exponential sum ${\cal A}_{p_N}^{(M,j)}(p_l)$, with $M=3$, $N=111547$ for the two different orders $j=2,3$, as functions of the parameter $p_{l} \equiv  l u_l$, expressed in units of $u_l$.
We can see that the two factors $l=331,337$, represented by stars, give complete constructive interference, despite  the other trial factors, represented by triangles, which present partially destructive interference.  All the other absolute maxima (represented by points) do not corresponding to integer factors for $N=111547$, but, in general, they are associated to the factors of different numbers $N^{\prime}=\alpha N$, if we use the apposite correspondent units $\frac{u_l}{\alpha}$ for
the parameter $p_l$. We can also note, as expected, that the peaks associated to
 the absolute maxima, in the case $j=3$, are sharper than the respective peaks, in the case $j=2$.
  On the other hand, increasing the order $j$, increase the value of the maxima of second order
  in the interference pattern. \label{contexp}}%
\label{sample-figure}
\end{center}
\end{figure}

\section{Continuous representation of exponential sums for factorizing different numbers}

The truncated exponential sum, in Eq. (\ref{exp}), can be straightforwardly extended to a continuous representation, in which the variable $l$ is a positive real number. This corresponds to considering a physical system, which is able to reproduce an interference pattern, given by the modulo squared of a truncated exponential sum, as a continuous function of values of the physical parameter $p_l$ in Eq. (\ref{cond2}), associated to a defined physical observable. This would allow us to get information about all the trial factors in a single measurement of the entire interference pattern, differently from the discrete approach in which independent measurements, for each trial factor, are necessary.

Fig.\ref{contexp} represents a continuous truncated exponential sum, with $M=3$, $N=111547$, for two different orders $j=2,3$. We can see that the two factors $l=331,337$ (represented by stars) give complete constructive interference. On the other hand, for the other trial factors (represented by triangles), there is partially destructive interference. Moreover, there are absolute maxima (represented by points) which do not correspond to integer trial factors.  In the next sections we will show how such maxima together with the whole continuous spectrum associated to these generalized exponential sums,  will turn out to be a powerful tool for distinguishing factors from non factors and factorizing different numbers at the same time. On the other hand, W\"{o}lk et al. have showed, in this issue, that the peaks of truncated Gauss sums at rational arguments give information about the factors, even if they do not correspond to integer trial factors \cite{wolk}.

\subsection{Distinction between factors and ``ghost" factors}
In the usual discrete representation of truncated exponential sums, in Eq. \ref{exp}, an important role is given
by the so called ``ghost" factors,  non factors which correspond to
values of the modulo squared of the exponential sums larger than the threshold
value of $\frac{1}{\sqrt{2}}$ \cite{ghost,expsum}.  In fact, in such an approach, in order to check that a ghost factor is not actually a factor, we must experimentally resolve the difference between the corresponding value of the modulo squared of the Gauss sum and the unitary value associated with the factors, for each independent measurement associated with each trial factor. Consequently a ghost factor  may not, in general, be distinguished from a real factor, if we take into account the variation of the values of the truncated exponential sum due to the experimental error.  In fact, such error, associated with the source, the instability of the physical system, and the detection, affects independently  each experimental measurement, obtained for each trial factor. For this reason, in the discrete approach, it is necessary to suppress the ghost factors, increasing the number of interfering terms in the sum. The threshold under which it is necessary to suppress the ghost factors and consequently the number $M$ of interfering paths increases with the experimental inefficiency\footnote{In particular, in the Ref. \cite{expsum}, it is shown that  at least $M\sim\sqrt[2j]{N}$ terms in the sum are necessary in order to suppress all the ghost factors under the threshold value of $\frac{1}{\sqrt{2}}$; however, in general, the necessary threshold can be larger than  $\frac{1}{\sqrt{2}}$.}.
A continuous representation of exponential sums, instead, allows us to check that ghost factors are effectively not factors, exploiting the whole continuous interference pattern. In such a pattern, it is easy to recognize that there is not total constructive interference associated with a ghost factor. In fact, the effective first order local maxima (total constructive interference) correspond to a non integer multiple of the unit $u$ (not a trial factor), in the neighborhood of the ghost factor we are considering. Such a local interference behavior is not affected by the experimental error, mentioned previously. Such error, in fact, affects only the signal to noise ratio and the width of the interference peaks of the entire interference pattern. It does not affect the rational values of $l$ corresponding to the effective local first order maxima and the symmetric behavior of the corresponding peaks, since the whole continuous pattern, different from the discrete approach, is obtained in a single interference measurement.\footnote{A detailed analysis of the experimental error associated to the reproduction of continuous exponential sums goes beyond the purpose of this paper and it will be described in a further publication.}.
Consequently, in the continuous approach, it is not necessary to experimentally resolve the values associated to the ghost factors respect to the unitary value corresponding to a factor. In fact, it is possible to distinguish between a ghost factor and the value associated with the effective local maxima, exploiting the symmetry of each peak in the continuous interference pattern and using a suitable resolution in the parameter $p_l$. In the discrete approach, instead, no matter how good the resolution in the parameter $p_l$, we have no clue if there is an effective maxima (not corresponding to a trial factor) in the neighborhood of the ghost factor. This is the reason, in the discrete representation, any independent indetermination associated with each single measurement of the Gauss sum, for each trial factor, compromises the distinction between factors and ghost factors. In conclusion, the continuous interference pattern, differently from the discrete approach, allows us to check that ghost factors do not correspond to maxima, even if the corresponding value of the modulo squared of the Gauss sum is close to unity. Consequently, suppression of ghost factors is not required, allowing a reduction of the number of interfering terms necessary to distinguish factors from non factors, compared to the discrete approach.  This implies a considerable reduction in the number of experimental resources, especially for factorizing larger numbers. In fact, the distinction between factors and ghost factors, depends only on the resolution in the parameter $p_l$, and the necessary resolution does not depend on the number of interfering paths.

In Fig.\ref{contexp}, for example, looking at the continuous interference spectrum, we can recognize, for both the case $j=2,3$, that $l=330,332,333,334,335,336$ are not factors (i.e. they do not correspond to absolute maxima), even if the sum assumes a value pretty close to unity. In particular, in their neighborhood, it is always possible to identify non integer values corresponding to the effective absolute maxima. We have showed, in Ref. \cite{prl}, that such a continuous interference pattern is indeed preserved in the actual experimental realization.

We will now compare the interference pattern of the truncated continuous exponential sums of two different orders  $j=2,3$, in Fig. \ref{contexp}.
It turns out, as expected, that, as the order $j$ of the exponential sum increases, the peaks associated with the absolute maxima becomes sharper. Consequently it is easier to check the relative difference between the value associated to the eventual "ghost" factors and the correspondent absolute maxima in their neighborhood. On the other hand, increasing the order $j$, increase the value of the second order maxima in the interference pattern. In order to suppress such maxima it is necessary to increase the number of terms $M$ in the sum.

 \subsection{Factorizing different numbers in a single run}
Now we want  to show that the continuous truncated exponential sum in Eq. (\ref{expphys}), as a function of the continuous parameter $p_l$, obtained in a single run of a definite experimental procedure, can be used to factorize not only $N$, but a generic number $N^{\prime}=\alpha N$, with $\alpha$ an apposite positive real number. In order to achieve this goal, we can apply the rescaling procedure introduced by Merkel et al. \cite{merkel2}, by simply rescaling the physical parameter $p_l$, defined in Eq. \ref{cond2}, to the value  $p_{l} \equiv  l \frac{u_l}{\alpha}$.

The trial factors correspond now to the discrete subset of values of $p_{l}$, with step  $\frac{u_l}{\alpha}$, where $\alpha$ defines the number $N^{\prime}$ we want to factorize. Obviously, for $\alpha=1$, $N^{\prime}$ coincides with $N$.
Such a rescaling procedure allows us to exploit the same experimental result, without running the experiment again. For example, we can exploit the continuous truncated sums for $N=111547$, in Fig. \ref{contexp}, for factorizing a different number $N^{\prime}=113230$. It turns out that the two factors $l=335,338$ correspond, respectively, to the two values  $p_{l} \equiv  330.021 u_l, 332.976 u_l$. In general, it is important to point out that all the absolute maxima of the continuous truncated exponential sum ${\cal A}_{p_N}^{(M,j)}(p_l)$, in Eq. \ref{expphys}, correspond to factors of definite numbers $N^{\prime}=\alpha N$ different from
$N$, if we use the apposite correspondent units $\frac{u_l}{\alpha}$ for
the parameter $p_l$.

\section{Implementation using an optical interferometer with variable optical paths}

\begin{figure}[t]
   \begin{center}
   \begin{tabular}{c}
   \includegraphics[width=7cm]{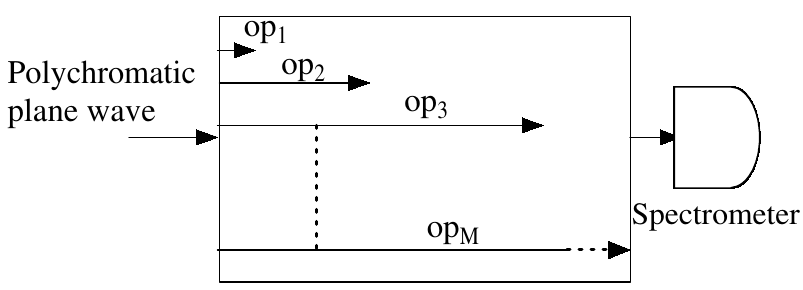}
   \end{tabular}
   \end{center}
\caption {Theoretical model of a generic $M$-path optical
interferometer, where $op_{m} \equiv  m^{j} N u$ is the length of
the $m_{th}$ optical path with $m=1,2,...,M$. The optical paths are
represented by arrows, which length increases quadratically respect to $m$, but the scheme is generalizable to orders $j>2$. The incoming electromagnetic field is
given by a polychromatic plane wave and the spectrum of the outgoing
field, given by the interference of all the $M$ optical paths, is
measured by a spectrometer\cite{prl}.\label{interf}}
   \end{figure}

A good physical system, for the implementation of the algorithm described so far, is given by a generic $M$-path optical interferometer (see Fig.
\ref{interf}),  with input signal given by an incoming polychromatic
 plane wave of intensity $|E_{in}(\lambda)|^2$, for each Fourier mode $\lambda$\cite{prl}.
The output signal is given by the coherent superposition of all the electromagnetic modes associated with the $M$ different optical paths and all the Fourier modes $\lambda$ associated with the source.
Using a spectrometer, we can measure the output electromagnetic intensity as a function of the wavelength $\lambda$:
\begin{eqnarray}\label{field}
|E_{out}(\lambda)|^2= \sum_{m=1}^{M+1} \exp\left[
2\pi  i \frac{op_{m}}{\lambda}\right],
\end{eqnarray}
where $op_{m}$ is the $m^{th}$ optical path in the interferometer.
The expression in Eq. (\ref{field}) corresponds to the modulo squared of the exponential sum in Eq. \ref{expphys}, where the physical parameters, in Eqs (\ref{cond2}) and (\ref{cond3}), are given by:

\begin{eqnarray}\label{pl}
p_{l}\equiv \lambda\equiv  l u_l,
\end{eqnarray}
\begin{eqnarray}\label{pNm}
 p_{N,m}\equiv op_{m}\equiv  (m-1)^j N u_N,
\end{eqnarray}
respectively, where we have taken into account that $u_N\equiv u_l$.
In this way, it is possible to find the factors, corresponding to the wavelengths which give maxima in the intensity spectrum, in a single run of the algorithm/experiment. In fact, it is important to point out that the measurement of the interference pattern is a single run interference measurement, obtained for example by a CCD camera; it is not a spectrum measurement obtained by scanning each single wavelength. Obviously, depending on the bandwidth of the optical devices, we need to measure the interference pattern for a certain number of different wavelength ranges. This simply means that the factoring analogue computer needs to include different optical interferometers, with associated optical devices suitable for each different wavelength range.

It is possible to apply the rescaling procedure described in the last section to this physical system. This means that the same output diffraction pattern obtained for factorizing a number $N$ can be used to infer the factors of different numbers $N^{\prime}$, simply by rescaling the wavelengths.

%%%%%%%%%%%%%%%%%%%%%%
%%%%%%%%%%%%%%%%%%%%%%

\subsection{Two possible interferometers}
We have seen in the last section that, in order to reproduce exponential sums with an $M$-path optical interferometer, the two conditions, in Eqs. \ref{pl} and \ref{pNm}, need to be satisfied.
The first condition is achieved by exploiting the entire spectrum of the polychromatic
incoming plane wave, which allows the reproduction of all the possible trial factors
at the same time.
On the other hand, the second condition is strictly related to the actual realization of the $M$-path optical interferometer.
In order to satisfy such a condition
we need to be able to manipulate either the indexes
of refraction $n_{m}$ or the lengths $d_{m}$, associated with the $M$ optical paths $op_{m}\doteq n_{m} d_{m}$, with $m=1,...,M$.
The first approach can be implemented by using a liquid crystal
grating. In the second approach, instead, we introduce a generalized
symmetric Michelson interferometer.

\subsubsection{Liquid crystal grating}
Let us analyze the first approach. First, we will describe an
interesting property which makes liquid crystals able to reproduce truncated
exponential sums. When we apply a variable voltage $V$ to a liquid
crystal cell interacting with an incoming plane wave, we can
observe a well defined dependence of the birefringence  of the liquid crystal on
the applied voltage\cite{lq}.  Such a definite behavior turns out to be a good
tool in order to reproduce the terms $(m-1)^{j}$, with $m=1,2,...,M$, in
Eq. (\ref{pNm}).
%%%%%%%%%%%%%%%%%%%%%%
\begin{figure}[t]
   \begin{center}
   \begin{tabular}{c}
   \includegraphics[width=9cm]{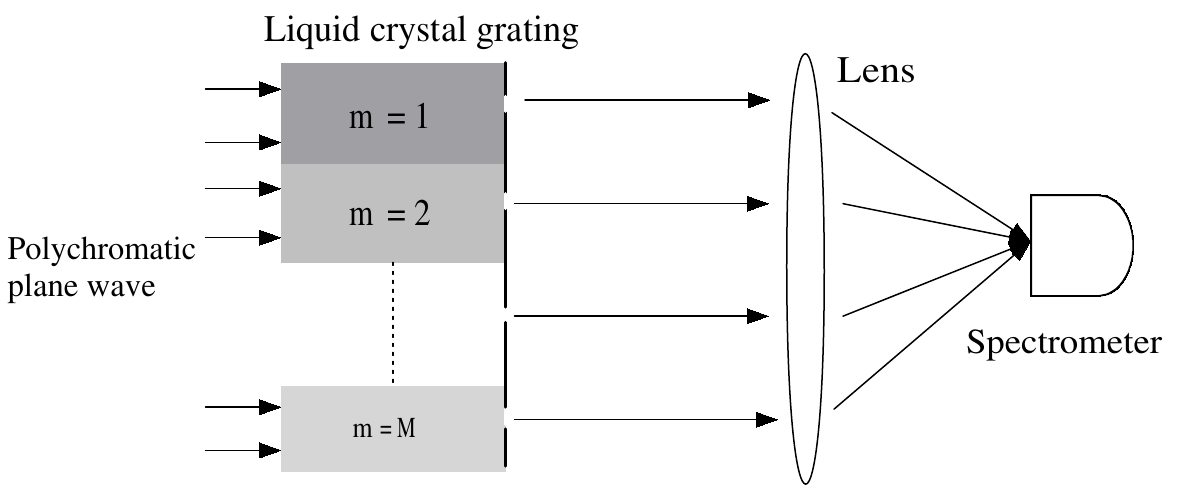}
   \end{tabular}
   \end{center}
\caption {Liquid crystal interferometer: a polychromatic plane wave,
in the ordinary mode, interacts first with a liquid crystal grating,
with $M$ regions and respective slits, and at the end with a lens. A
spectrometer measures the intensity of the light as a function of
the wavelength in the focal plane of the lens\cite{prl}.\label{grating}}
   \end{figure}
%%%%%%%%%%%%%%%%%%%%%%

The $M$ terms in the truncated exponential
 sum  correspond, respectively, to $M$ different
  regions in a liquid crystal cell with the same thickness $d_{m}\equiv d$.

The basic experimental setup is shown in Fig. \ref{grating}. An
incoming polychromatic plane wave, in the ordinary mode, interacts
with a liquid crystal grating with $M$ slits. Such a grating
consists of $M$ liquid crystal regions, with $M$ different variable
applied voltages $V_{m}$, where $m=1,2,...,M$, and a slit at the end
of each region.

 So, when the
incoming polychromatic plane wave interacts with the liquid crystal
grating, it gives rise to $M$ different electromagnetic phase terms,
which can be manipulated in an appropriate way, by varying the applied
voltages $V_{m}$ until the condition in Eq. (\ref{pNm}) is satisfied. Such
terms superpose
coherently in the focal point
 of a lens, reproducing a continuous truncated exponential sum.

Of course, in the experimental realization of such an approach, we need
to take into account the dispersion associated with the broadband
spectrum of the source. Such problem can be overcome by performing
several measurements in different ranges of the spectrum of the
light source such that the relative dispersion in each range is
negligible.

It is also important to point out that the larger the maximum
achievable optical path, the larger is also the maximum achievable
truncation parameter $M$ in Eq. (\ref{expphys}).
 Unfortunately, in the liquid crystal approach, the maximum range of variation
  of the optical paths is limited by the thickness $d$ of the liquid crystal cells and by the birefringence,
   calculated when no voltage is applied.
 Consequently,  both these parameters determine the maximum number of terms in the truncated
exponential sum and the maximum range of possible numbers $N$ we can
factorize.

\subsubsection{Generalized symmetric Michelson interferometer}
We have seen one possible way of varying the optical paths, in an
$M$-path interferometer, in order to obtain phase terms of order
$j$. We will now describe how to achieve the same result by varying, in free
space, the path lengths $d_{m}$ in Eq. (\ref{pNm}). In this case, we
do not encounter any
 problem associated with dispersion.

\begin{figure}[t]
   \begin{center}
   \begin{tabular}{c}
   \includegraphics[width=9cm]{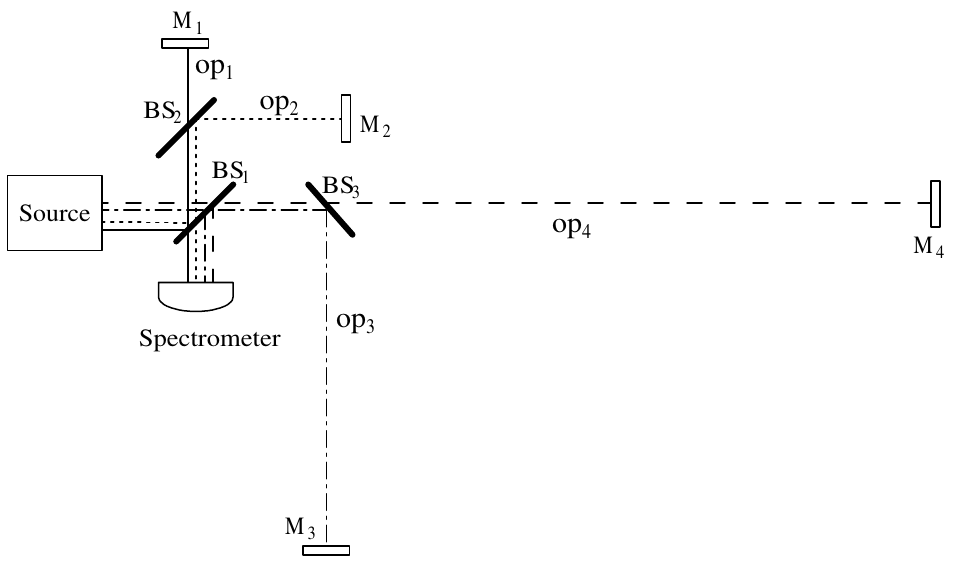}
   \end{tabular}
   \end{center}
\caption {Generalized symmetric $M$-path Michelson interferometer
for the realization of exponential sum with truncation parameter
$M=4$\cite{prl}. The usual two-paths Michelson interferometer is generalized
to an $M$-path interferometer, using $M-1$ beam splitters. The $M=4$
interfering optical paths, indicated with dashed, dashed-dotted,
continuous, and dotted lines can be varied by moving longitudinally
the mirrors $M_{1}$, $M_{2}$,
 $M_{3}$ and $M_{4}$, respectively, in order to satisfy the condition (\ref{pNm}). \label{miche}}
   \end{figure}

 This approach can be implemented
 exploiting the multi-path interference in a generalized symmetric
  Michelson interferometer in free space. The usual two-path Michelson
interferometer is generalized, in a symmetric way, to an $M$-path
interferometer, using $M-1$ beam splitters.  In Fig. (\ref{miche})
we have represented, for simplicity, the case $M=4$ (obviously such
 an approach can be extended to a generic $M$). In this case, the
four interfering optical paths can be varied arbitrarily by
translating the mirrors $M_{1}$, $M_{2}$,
 $M_{3}$ and $M_{4}$, respectively. Our factorization algorithm
   can be easily implemented using such an interferometer, giving all the factors of any number $N$ in a single run. Moreover, because the
 lengths of the interfering paths can be in principle as large as we want,
 there is no limit to
 the maximum achievable truncation parameter $M$ and
 order $j$ of the exponential sum we want to reproduce. This is
 another aspect in favor of this approach, rather than the one based
 on a liquid crystal grating.

 %%%%%%%%%%%%%%%%%%%%%%
%%%%%%%%%%%%%%%%%%%%%%

\section{Conclusions}

We have described a generic factorization algorithm based on the continuous representation of Gauss sums, generalizable to orders $j>2$.
Such an algorithm allows us, for the first time, to find all the factors of a number $N$ in a single run without precalculating the ratio $N/l$. Moreover we have shown that, using a rescaling procedure, it is possible to factorize different numbers $N^{\prime}$ analyzing the same output interference pattern.

The continuous generalization of the Gauss sums approach allows us to verify that a ghost factor does not correspond to total constructive interference in the continuous interference pattern. In fact, it is possible to identify  the  effective position of the interference maximum, by looking at the local neighborhood of the considered ghost factor. This allows us to substantially reduce the number of interfering paths with respect to the  the discrete approach, with a consequent reduction in the number of experimental resources, especially for factorizing larger numbers.

  It is also
 possible to use this algorithm for the realization of truncated
exponential sums with a number of terms $M'<M$, randomly chosen
among the total $M$ terms in Eq.
\ref{exp}\cite{randomweber,randompeng}, in order to further reduce the
number of resources.

We have also introduced two possible $M$-path optical interferometers, which can be used to experimentally realize  this algorithm: a liquid crystal grating and a generalized symmetric Michelson interferometer. An experimental proof of principle of this algorithm has indeed been realized using an $M=3$-path Michelson interferometer\cite{prl}.

\vspace{6mm}

\hspace{0mm}{\large\bf  Acknowledgements}

\vspace{3mm}

A. Rangelov  proposed, independently from us, an interesting factorization scheme based on exponential sums, which also overcomes pre-calculation of the ratio $N/l$ present in the past proposals. Rangelov's approach is based on the use of a Mach-Zehnder interferometer\cite{rangelov}, with fixed indexes of refraction in each interfering path. This unfortunately means that we need to change the optical setup, depending on the number we want to factorize.
Moreover such a scheme does not include a way to check all the trial factors at the same time and for different values of $N$.

The authors thank M. D'Angelo, J. Franson, T. Pittman, A. Rangelov, M. H. Rubin,
G. Scarcelli, S. W\"{o}lk, T. Worchesky and, in particular, W. Schleich for
useful suggestions and stimulating discussions.

%%%%%%%%%%%%%%%%%%%%%%%%%%%%%%%%%%%%

\label{lastpage}

\end{document}